\documentclass[12pt]{article}
\usepackage{amsfonts,amstext}

\input epsf

\let\a=\alpha \let\b=\beta \let\g=\gamma \let\d=\delta \let\e=\varepsilon

\let\s=\sigma \let\t=\tau  

\let\w=\omega \let\G=\Gamma

\let\pa=\partial
\def\be{\begin{equation}}
\def\ee{\end{equation}}
\def\ba{\begin{array}}
\def\ea{\end{array}}

\def\dalemb#1#2{{\vbox{\hrule height .#2pt
        \hbox{\vrule width.#2pt height#1pt \kern#1pt
                \vrule width.#2pt}
        \hrule height.#2pt}}}

\newcommand{\bea}{\begin{eqnarray}}
\newcommand{\eea}{\end{eqnarray}}

\def\ep{{\epsilon}}
\def\vep{{\varepsilon}}
\def\bR{{{\Bbb R}}}

\def\bZ{{{\Bbb Z}}}

\def\Ub{{\widetilde U}}
\def\Nb{{\widetilde N}}

\def\xb{{\tilde x}}
\def\ab{{\tilde a}}

\def\bb{{\tilde \beta}}

\newcommand{\preprint}[1]{\begin{table}[t]  
             \begin{flushright}               
             {#1}                             
             \end{flushright}                 
             \end{table}}                     
\renewcommand{\title}[1]{\vbox{\center\LARGE{#1}}\vspace{5mm}}
\renewcommand{\author}[1]{\vbox{\center#1}\vspace{5mm}}
\newcommand{\address}[1]{\vbox{\center\em#1}}
\newcommand{\email}[1]{\vbox{\center\tt#1}\vspace{5mm}}

\begin{document}

\begin{titlepage}
\preprint{hep-th/0610261 \\
DAMTP-2006-100 \\
NSF-KITP-06-96}

\title{Einstein-Maxwell gravitational instantons and \\
five dimensional solitonic strings}

\author{Maciej Dunajski${}^1$ and Sean A. Hartnoll${}^2$}

\address{${}^1$ DAMTP, Centre for Mathematical Sciences,
Cambridge University\\
Wilberforce Road, Cambridge CB3 OWA, UK \\
\vspace{0.3cm}
${}^2$ KITP, University of California Santa Barbara \\
 CA 93106, USA}

\email{m.dunajski@damtp.cam.ac.uk, hartnoll@kitp.ucsb.edu}

\abstract{
We study various aspects of four dimensional Einstein-Maxwell
multicentred gravitational instantons. These are half-BPS
Riemannian backgrounds of minimal ${\mathcal{N}}=2$ supergravity,
asymptotic to $\bR^4, \bR^3\times S^1$ or $AdS_2\times S^2$.
Unlike for the Gibbons-Hawking solutions, the topology is not
restricted by boundary conditions. We discuss the classical metric
on the instanton moduli space. One class of these solutions may be
lifted to causal and regular multi `solitonic
strings', without horizons, of 4+1 dimensional ${\mathcal{N}}=2$ supergravity,
carrying null momentum.}

\end{titlepage}

\tableofcontents

\section{Introduction}

\subsection{Motivation}

Instantons play a key role in the nonperturbative dynamics of
Yang-Mills theories, and indeed in a wide range of quantum mechanical
systems. One useful property of instantons is that they can allow a
semiclassical description where a full treatment is either difficult
or even ill defined, as in the case of gravity. At the other extreme,
in supersymmetric theories instantons are crucial in obtaining
exact results.

Within the programme of Euclidean Quantum Gravity,
multicentred gravitational instantons followed hotly on the tails of
their Yang-Mills counterparts \cite{Hawking:1976jb,
  Gibbons:1979zt}. However, while the Gibbons-Hawking metrics have found a
surprising range of physical applications, their dynamical role within
quantum gravity remains unclear. One reason for this is that if the
instanton contains more than one centre, it is no longer
Asymptotically Euclidean ($\sim \bR^4$) or Asymptotically Flat ($\sim
\bR^3 \times S^1$). These are the
most natural asymptotics for infinite volume quantum gravity at
zero or finite temperature, respectively. In contrast, at constant
large radius the multicentred Gibbons-Hawking spaces tend to $S^1$
fibred over $S^2$ with increasingly high Chern number.

Said differently, the boundary conditions determine the
gravitational instanton topology. There is no sum over different
spacetime topologies for a fixed asymptotics. In this sense, the
Gibbons-Hawking spaces do not provide a semiclassical realisation
of spacetime foam.

It is therefore of interest to study gravitational theories in
which arbitrarily high instanton number is allowed with fixed
asymptotics. One example of such a theory is conformal gravity, in
which the Einstein-Hilbert term is replaced by the Weyl curvature
squared \cite{Strominger:1984zy,L91,Hartnoll:2004rv}. Despite some
rather attractive features of the gravitational instantons in this
theory, the physical status of the theory itself is uncertain due
to problems with higher derivative Lagrangians and unitarity.

In this paper we emphasise that Einstein-Maxwell theory also
admits regular multicentred instantons with arbitrarily
complicated topology for fixed asymptotics. These solutions have
essentially appeared before in the literature
\cite{Whitt:1984wk,Yuille:1987vw}. Various unsatisfactory
aspects of these previous treatments, for instance we have
preferred to use a Riemmanian Maxwell field that is real, have
lead us to carry out a systematic study {\it de novo}. We
furthermore extend our understanding of Einstein-Maxwell
gravitational instantons through discussions of uniqueness,
supersymmetry, moduli space metrics and lifts to five dimensions.
This final point may be of independent interest.

\subsection{Summary}

In Section 2 we present the instanton solutions. We detail the
possible asymptotics: $\bR^4, \bR^3 \times S^1$ and $AdS_2 \times S^2$,
and local versions thereof. We show that the solutions are half-BPS
when embedded into minimal ${\mathcal N} = 2$ supergravity and that they are
all the regular Riemannian half-BPS solutions. Finally, we evaluate
the action of the solutions. The Asymptotically Euclidean case is
found to only be well defined when a certain linear combination of the
charge and potential is fixed at infinity.

In section 3 we discuss the moduli space metric on the
Einstein-Maxwell instantons. We consider in some detail the
ambiguities involved in finding an inner product on the space of
metric fields. We show that there is a preferred inner product
which is inherited from the action and for which zero modes are
orthogonal to pure gauge modes.

Section 4 shows how the four dimensional instantons may be lifted
to solitons of five dimensional Einstein-Maxwell-Chern-Simons
theory, or minimal ${\mathcal N} = 2$ supergravity in five
dimensions. Generically the lifted solutions are either singular
or contain closed timelike curves. However, we find that one class
of solutions lift to regular, causal plane fronted wave spacetimes
with the fields localised in lumps orthogonal to the wave
propation. We call these `solitonic strings' as they do not have
an event horizon.

Section 5 briefly discusses the slow motion of the five
dimensional solitons. Unlike in the case of the Gibbons-Hawking
instantons and their lift to Kaluza-Klein monopoles, it seems that
there is not a direct connection between the four dimensional
instanton moduli space metric and the five dimensional soliton
slow motion moduli space metric in our case.

We end with a discussion of possible physical applications of
these multicentred Einstein-Maxwell instantons, and directions for
future work.

\section{The gravitational instantons}

\subsection{The solutions}
\label{sec:metric}

The gravitational instantons on a four dimensional manifold $M_4$
are solutions to the Einstein-Maxwell equations with Riemannian
signature
\bea\label{eq:4deqns}
G_{a b} & = & 2 F_a{}^c F_{b c} - \frac{1}{2} g_{a b} F^{c d} F_{c d} \,, \\ \nonumber
\nabla_a F^{a b} & = & 0 \,.
\eea
The metric is given by
\be
\label{IWmetric}
g^{(4)} = \frac{1}{U \Ub} (d\t + {\bf \w})^2 + U \Ub d{\bf x}^2
\,,
\ee
where the functions $U,\Ub$ and the one form $\w$ depend on ${\bf
  x}=(x, y, z)$ and satisfy
\bea
\label{IWequations}
\nabla^2 U = \nabla^2 \Ub = 0 \,, \nonumber \\
\nabla \times \w = \Ub \nabla U - U \nabla \Ub \,.
\eea
We will work with four dimensional tangent space indices, $a,b,...$ and the vierbeins
\be
e^4 = \frac{1}{(U \Ub)^{1/2}} (d\t + {\bf \w}) \,, \qquad e^{i} =
(U \Ub)^{1/2} dx^i \,.
\ee
The electromagnetic field strength may now be written
\bea\label{eq:fieldstrength}
F_{4 i} & = & \frac{1}{2} \pa_i \left[U^{-1} - \Ub^{-1}
  \right] \,, \nonumber \\
F_{i j} & = & \frac{1}{2} \e_{ijk} \pa_k \left[U^{-1} + \Ub^{-1}
  \right] \,,
\eea
where the derivatives are partial derivatives with respect to the
corresponding spacetime indices. One can check that this field
strength satisfies the Bianchi identities, and thus locally at least
we can write $F=dA$. Our expressions for the
field strength in Riemannian signature differ slightly from others in
the literature \cite{Whitt:1984wk,Yuille:1987vw} which were not real.
In particular the Riemannian  Majumdar-Papapetrou metrics with
$U=\Ub$ have purely magnetic field strength $F=-2\star_3 dU$.

These backgrounds were first found in the Lorentzian regime by Israel
and Wilson \cite{israelwilson} and by Perj\'es \cite{perjes} as a
stationary generalisation of the static Majumdar-Papapetrou multi
black hole solutions. However, it was shown by Hartle and Hawking that
all the non static solutions suffered from naked singularities
\cite{hartlehawking,Chrusciel:2005ve}.

With Riemannian signature however, regular solutions exist
\cite{Whitt:1984wk,Yuille:1987vw}. We can take
\be\label{eq:sumpoles}
U = \frac{4\pi}{\b} + \sum_{m=1}^N \frac{a_m}{\mid {\bf x} - {\bf
x}_m \mid} \,,
\qquad \Ub = \frac{4\pi}{\bb} + \sum_{n=1}^\Nb \frac{\ab_n}{\mid {\bf x} -
  {\bf \xb}_n \mid} \,,
\ee
in these expressions $\b,\bb,a_m,{\bf x}_m,\ab_n,{\bf
\xb}_n,N,\Nb$ are constants. For the signature to remain
 $(+,+,+,+)$ throughout we can require $U,\Ub > 0$ which in turn
requires $a_m,\ab_n > 0$. From the explicit forms for $U$ and
$\Ub$ in (\ref{eq:sumpoles}) we can write down explicit
expressions for the one forms ${\bf \w}$ and $A$, which so
far we have only defined implicitly. These are given in Appendix
A.

If there is at least one non coincident centre, ${\bf x}_m \neq
{\bf \xb}_n$, regularity requires that $\t$ is identified with
period $4\pi$ and that the constants satisfy the following
constraints at all the non-coincident centres
\be \label{eq:constraints}
U({\bf \xb}_n) \ab_n = 1 \,, \qquad \Ub({\bf x}_m) a_m = 1 \,, \qquad \forall
m,n \,.
\ee
Given the locations of the centres $\{ {\bf x}_n,{\bf \xb}_m\}$,
these constraints may be solved uniquely for the $\{a_n,\ab_m\}$
\cite{Yuille:1987vw}. When $\frac{4\pi}{\b} =
\frac{4\pi}{\bb} = 0$ the solution is only unique up to the overall
scaling
\be\label{eq:scaling}
U \to e^{s} U \,, \qquad \Ub \to e^{-s} \Ub\,.
\ee
In general this scaling leaves the metric invariant and induces a linear duality
transformation on the Maxwell field mapping solutions to solutions
\be
{\bf E} \to \cosh s \, {\bf E} + \sinh s \, {\bf B} \,, \qquad
{\bf B} \to \sinh s \, {\bf E} + \cosh s \, {\bf B} \,.
\ee
The rescaling does not leave the action and other properties of the
solutions invariant.

The constants $\b$ and $\bb$ determine the asymptotics of the
solution. There are three possibilities:

\begin{itemize}
\item The case $\frac{4\pi}{\b} = \frac{4\pi}{\bb} \neq 0$
gives an Asymptotically Locally Flat metric, tending to
an $S^1$ bundle over $S^2$ at infinity, with first Chern number $N-\Nb$.
Without loss of generality we have rescaled the harmonic functions
using (\ref{eq:scaling}) so that $\b = \bb$.
Equations (\ref{eq:constraints}) now imply that $\sum a_m - N = \sum
\tilde{a}_n - \Nb$. If $N=\Nb$ the asymptotic bundle is trivial
and we obtain Asymptotically Flat ($\sim \bR^3 \times S^1$) solutions.

\item The case $\frac{4\pi}{\b} = 0$, $\frac{4\pi}{\bb} = 1$
gives an Asymptotically Locally Euclidean metric, tending to $\bR^4 /
\bZ_{|N-\Nb|}$. We have used the rescaling (\ref{eq:scaling})
to set $\frac{4\pi}{\bb} = 1$ without loss of generality.
In this case the constraints (\ref{eq:constraints}) require that $\sum a_m =
N - \Nb$. Of course we can reverse the roles of $\b$ and $\bb$. If $N
= \Nb + 1$ the solution is Asymptotically Euclidean ($\sim \bR^4$).

\item The case $\frac{4\pi}{\b} = \frac{4\pi}{\bb} = 0$ leads to
an Asymptotically Locally Robinson-Bertotti metric, tending to $AdS_2
\times S^2$ or $AdS_2/\bZ \times S^2$. The former case only arises
if all of the centres are coincident, so that $U = \Ub$, and $\t$ need
not be made periodic. For both these
asymptotics, the constraints (\ref{eq:constraints}) require that $N =
\Nb$. We may further use the rescaling (\ref{eq:scaling}) to set
$\sum a_m = \sum \tilde{a}_n$.
\end{itemize}

As Riemannian solutions, the backgrounds are naturally thought of as
generalisations of the Gibbons-Hawking multicentre metrics which in
fact they include as the special case $\Ub=1$, albeit with an
additional antiselfdual Maxwell field. A crucial new aspect of the
Asymptotically Locally Euclidean (ALE) and Asymptotically Locally Flat (ALF)
Israel-Wilson-Perj\'es solutions is that when
\be
N = \Nb \pm 1 \; \text{(for ALE)} \qquad \text{or} \qquad N = \Nb \; \text{(for
ALF)} \,,
\ee
the fibration of the $\tau$ circle over $S^2$ at infinity is
trivial and the metrics do not require the $\bZ_N$ identifications
at infinity that are needed in the Gibbons-Hawking case. The
spacetimes are therefore strictly Asymptotically Euclidean and
Asymptotically Flat respectively in these cases. The Euler number
is given by $\chi = N + \Nb-1$ in the ALF and ALE cases
\cite{Yuille:1987vw}. Thus the spaces
admit arbitrarily complicated topology, not restricted by the
asymptotic topology, and provide a semiclassical realisation of
spacetime foam in quantum Einstein-Maxwell theory.

The metric (\ref{IWmetric}) has vanishing scalar curvature. If
$U$ or $\Ub$ is constant then (\ref{IWmetric}) is Ricci flat,
and hyperK\"ahler. It is natural to ask whether any other
special choices of harmonic functions $U$ and $\Ub$ lead
to scalar flat K\"ahler metrics with a symmetry $\pa/\pa \t$
preserving the K\"ahler structure. Such metrics would be
conformally anti--self--dual and thus interesting in twistor theory.
The answer is negative.
From \cite{L91} any such metric is of the
form
\be
g^{(4)}= \frac{1}{\cal W} (d\tau +\omega)^2 + {{\cal W}} h^{(3)}\,,
\ee
where the metric $h^{(3)}$ on the three dimensional orbit
space of $\pa/\pa \t$, and the function ${\cal W}$ on this space
satisfies a coupled nonlinear system of PDEs. In the case that
$h^{(3)}$ is flat the equations reduce to
\be
\nabla \times {\bf \w} =\nabla {\cal W}.
\ee
Therefore ${\cal W}=U\Ub$ is harmonic, and then
(\ref{IWequations}) implies that
$\Ub$ is a constant.

\subsection{Killing spinors}

The solutions have the further important property of admitting two
complex Killing spinors. These satisfy
\be\label{eq:susy}
e^{\mu}{}_a \pa_\mu \e + \frac{1}{4} \left[\w^{b c}{}_a\, \G_{b c} + i F^{b
    c}\, \G_{b c}\, \G_a  \right] \e = 0 \,,
\ee
where $\w^{b c}{}_a$ are the components of the  the spin connection
one form $\w^{bc}$
defined by $de^b = \w^{b
  c} \wedge e^{c}$. We use Greek letters $\mu,\nu,...$ to denote
Euclidean spacetime indices. Our gamma matrix conventions are
given in Appendix B, as is the spin connection for the background.
With these conventions one may solve the equation (\ref{eq:susy})
to find
\be\label{eq:spinor}
\e =
\left( \begin{array}{c}
U^{-1/2} \e_0 \\
i \Ub^{-1/2} \e_0
\end{array}
\right) \,,
\ee
where $\e_0$ is a constant two-component complex spinor: $\pa_\mu
\e_0 = 0$.

Within Einstein-Maxwell theory, the Killing spinors imply that the
solutions saturate a Bogomolny bound \cite{Gibbons:1982fy}. It is
also natural to view the solutions as half-BPS states of four
dimensional ${\mathcal{N}}=2$ supergravity
\cite{Ferrara:1976fu}. This theory has a complex
spin-$\frac{3}{2}$ Rarita-Schwinger field as well as the graviton
and photon. In fact, in a paper that anticipated current interest in
classifying supersymmetric solutions, Tod has shown that the
Lorentzian version of these solutions are all the supersymmetric
solutions to ${\mathcal{N}}=2$ supergravity with a timelike
Killing spinor \cite{Tod:1983pm}.

In the following subsection we shall repeat Tod's analysis in the
Riemannian case. As well as recovering the local form of the
metric, it will find that $|\nabla U^{-1}|$ and $|\nabla
\Ub^{-1}|$ are both bounded\footnote{This is stronger than
  the Lorentzian result of \cite{Chrusciel:2005ve} where the separate
  bounds cannot be established.}. Combined with a
result from analysis \cite{chrusciel_nad}, it will follow that (\ref{IWmetric}) together
with (\ref{eq:sumpoles}) is the most general regular supersymmetric
solution to minimal ${\mathcal{N}}=2$ supergravity. To put it differently, only
harmonic functions with a finite number of point sources lead to regular
metrics.

As usual, given Killing spinors $\e$ and $\eta$ we can build differential forms. In
particular, we have the one forms
\be
V = \frac{1}{2} \bar \eta \G_a \e \, e^a \,, \qquad K = \frac{1}{2}
\bar \eta \G_5 \G_a \e \, e^a \,,
\ee
and the two form
\be
\Omega = - \frac{i}{2} \bar \eta \G_{ab} \e \, e^a \wedge e^b \,.
\ee
In our representation of the Clifford algebra, given in the
appendix, all the gamma matrices are hermitian and therefore bar
simply denotes complex conjugation. From the Killing spinor
condition (\ref{eq:susy}) we have that
\be
d \Omega = - 2 V \wedge F \,, \qquad d V = 0 \,, \qquad \nabla_{(a}
K_{b)} = 0 \,.
\ee
With a little more work one can also show that
\bea\label{eq:domega}
\nabla_a \Omega_{bc} & = & 2 V_a F_{bc} - 4 F_{a [b} V_{c]} + 4 V^d
F_{d [b} g_{c] a} \,, \nonumber \\
\nabla_a V_b & = & \frac{1}{4} F^{cd} \Omega_{cd} g_{ab} + F^{c}{}_{(a}
\Omega_{b) c} \,.
\eea
In fact there is more structure. The two form $\Omega$ can be split
into self dual and anti-self dual parts: $\Omega = \Omega^+ +
\Omega^-$. One can then show that $\Omega^+$ and $\Omega^-$ separately
satisfy the first equation in (\ref{eq:domega}) with $F$ replaced
by its self dual, $F^+$, and anti-self dual, $F^-$, parts
respectively.

Three important cases giving real forms are when $\eta = \e = \e^I$, for $I =
1,2,3$, which are defined by $\e_0$ in
(\ref{eq:spinor}) satisfying $\bar \e_0^I \t^J \e_0^I = \delta^{I
  J}$ and $\bar \e_0^I \e_0^J = \delta^{I J}$. For these cases we find
\be
V^I = d x^I \,, \qquad K = \frac{1}{U \Ub} \left(d\t + {\bf \w} \right) \,,
\ee
and
\be
\label{OmegaI}
\Omega^I = \left(U^{-1} - {\tilde U}^{-1} \right) e^4 \wedge
e^I + \frac{1}{2} \left(U^{-1} + {\tilde U}^{-1} \right)
\epsilon^{I j k} e^j \wedge e^k \,.
\ee
Raising the index, the Killing vector is $K =
\pa/\pa \t$ as we should expect.

\subsection{Uniqueness of the solutions}
\label{killing_sp}

Here we show that the solution (\ref{IWmetric}),
(\ref{eq:fieldstrength}) with the harmonic functions described by
(\ref{eq:sumpoles}) and satisfying the constraints
(\ref{eq:constraints}) is the most general regular Einstein-Maxwell
instanton with a complex Killing spinor.

In this section it will be convenient to write the Dirac spinor
$\vep=(\alpha^A, \beta_{A'})$ as a pair of complex two-component
spinors. When dealing with these spinors we use the conventions
given in Appendix C. With positive signature, spinor conjugation
preserves the type of spinors. Thus if $\alpha_{A}=(p, q)$ we can
define $\hat{\alpha}_A=(\overline{q}, -\overline{p})$ so that
${\hat{\hat{\alpha}}}_A=-\alpha_A$. This hermitian conjugation
induces a positive inner product
\be
\alpha_A\hat{\alpha}^A=\ep_{AB}\alpha^B\hat{\alpha}^A=|p|^2+|q|^2 \,.
\ee
We define the inner product on the primed spinors in the same way.
Here $\ep_{AB}$ and $\ep_{A'B'}$  are covariantly constant symplectic
forms with $\ep_{01}=\ep_{0'1'}=1$. These are used to raise and lower spinor indices
according to $\alpha_B=\ep_{AB}\alpha^A, \alpha^B=\ep^{BA}\alpha_A$, and
similarly for primed spinors. In terms of our gamma matrices, $\hat \e
= \Gamma^{31} \bar \e$.

The Killing spinor equation (\ref{eq:susy}) becomes
\be
\label{spinor_cond}
\nabla_{AA'}\alpha_B-i \sqrt{2}\phi_{AB}\beta_{A'}=0\,,
\qquad
\nabla_{AA'}\beta_{B'}+i\sqrt{2}\tilde{\phi}_{A'B'}\alpha_{A}=0\,,
\ee
where the spinors $\phi$ and $\tilde{\phi}$ are symmetric in their respective
indices and give the anti-self dual and self dual parts of the electromagnetic
field
\be
F_{ab}=\phi_{AB}\ep_{A'B'}+\tilde{\phi}_{A'B'}\ep_{AB}\,.
\ee
Suppose that $\e = (\alpha^A, \beta_{A'})$ solves the Killing spinor
equation (\ref{spinor_cond}). Now we can reconstruct the spacetime
metric and Maxwell field.

Define
\be
\label{definitions_UU}
U=(\alpha_A\hat{\alpha}^A)^{-1}, \qquad
\Ub=(\beta_{A'}\hat{\beta}^{A'})^{-1}.
\ee
In our positive definite case, these two inverted functions do not
vanish unless $\a$ or $\b$ vanish.
In the Lorentzian case their possible vanishing leads
to plane wave spacetimes \cite{Tod:1983pm}. If $\a$ or $\b$ vanish
identically, we recover the Gibbons-Hawking solutions.
Now define a (complex) null tetrad
\be
X_a=\alpha_A\beta_{A'}, \qquad \overline{X}_a=\hat{\alpha}_A\hat{\beta}_{A'},
\qquad Y_a=\alpha_A\hat{\beta}_{A'},\qquad \overline{Y}_a=-\hat{\alpha}_A{\beta}_{A'}\,.
\ee
We can check that $\hat \e$ is also a solution to the Killing spinor
equation (\ref{spinor_cond}).
It therefore follows from (\ref{spinor_cond}) that  $X_a,  \overline{X}_a,
Y_a-\overline{Y}_a$  are gradients and that
$K_a= Y_a+\overline{Y}_a$ is a Killing vector.
Now define local coordinates
$(x, y, z, \tau)$ by
\be
X=\frac{1}{\sqrt{2}}(dx+idy), \qquad (Y - \overline{Y})=
i \sqrt{2}dz, \qquad K^{a}\nabla_a=\sqrt{2}\frac{\partial}{\partial\tau}\,,
\ee
where the form $X = X_a e^a = X_{A A'} e^{A A'}$ and similarly for
$Y,\overline{Y}$. The vector $K$ Lie derives
the spinors $(\alpha_A, \beta_{A'})$, implying that $U$ and $\Ub$ are
independent of $\t$.

The metric is now given by $ds^2 = \ep_{A B} \ep_{A' B'} e^{A A'} e^{B
B'}$. This expression may be evaluated by noting that from
(\ref{definitions_UU}) we have $\ep_{AB} = U (\a_A \hat \a_B - \a_B
\hat \a_A)$ and similarly for $\ep_{A'B'}$. Using the fact that from
the above definitions $K_aK^a=2(U\Ub)^{-1}$, we find that the metric takes the form
(\ref{IWmetric}) for some one form ${\bf \omega}$. The next step is to
find ${\bf \omega}$.

The definitions of $U,\Ub$ and $K$ together with (\ref{spinor_cond}) imply
\be
\label{Killing_relation}
\nabla_a K_b= i \sqrt{2} \left[ {\Ub}^{-1}\phi_{AB}\ep_{A'B'} +
U^{-1}\tilde{\phi}_{A'B'}\ep_{AB} \right] \,,
\ee
and
\be
\label{U_phi_relations}
\nabla_a U^{-1}= i\sqrt{2}\phi_{AB}K^{B}_{A'}, \qquad
\nabla_a \Ub^{-1}= -i\sqrt{2}\tilde{\phi}_{A'B'}K_{A}^{B'}\,.
\ee
The formulae in (\ref{U_phi_relations}) may be inverted to find
expressions for $\phi_{AB}$ and $\tilde \phi_{A'B'}$, using $K_B^{A'} K^{B C'}
= \frac{1}{2} \ep^{A' C'} K_{D E'} K^{D E'}$. Substituting the result into
(\ref{Killing_relation}) yields the expression (\ref{IWequations}) for
$\nabla\times {\bf \omega}$.

Finally, differentiating the relations
(\ref{spinor_cond}) shows that the energy momentum tensor
is that of Einstein--Maxwell theory:
$T_{ab}=2\phi_{AB}\tilde{\phi}_{A'B'}$. The Maxwell equations
\be
\nabla^{AA'}\phi_{AB}=0, \qquad \nabla^{AA'}\phi_{A'B'}=0 \,,
\ee
now imply that $U$ and $\Ub$ are harmonic on $\bR^3$. This completes
the local reconstruction of the solution from the Killing spinors.

So far everything has proceeded as in \cite{Tod:1983pm} with minor differences
in the reality conditions. The main difference arises in global regularity
considerations which lead us to consider the invariant
\begin{eqnarray}
F_{ab}F^{ab}&=&2(\phi_{AB}\phi^{AB}+\tilde{\phi}_{A'B'}\tilde{\phi}^{A'B'})
\nonumber\\
&=& |\nabla U^{-1}|^2+|\nabla \Ub^{-1}|^2,
\end{eqnarray}
where the norm of the gradients is taken with respect to the flat
metric on $\bR^3$, and we have used (\ref{U_phi_relations}).
Regularity requires this invariant
be bounded. Therefore both $|\nabla U^{-1}|$ and $|\nabla \Ub^{-1}|$
must be bounded. The various boundary conditions we have described imply that
$U$ and $\Ub$ are regular as $|\bf x|\rightarrow \infty$. In particular,
they are both regular outside a ball $B_R$ of sufficiently large
radius $R$ in $\bR^3$.

The coordinates $\{ {\bf x}, \tau\}$ cover $\bR\times
(\bR^3 \setminus {\cal S})$, where ${\cal S}$ is the compact subset of $B_R$ on which
$U$ or $\Ub$ blow up.
A theorem from \cite{chrusciel_nad} can now be applied separately to
both harmonic functions to prove that ${\cal S}$ consists of a finite
number of points. In fact
\be
\#{\cal S}<\mbox{max}\{|\nabla U^{-1}|,  |\nabla \Ub^{-1}|\} |U(p)
+\Ub(p)|\;R+1,
\ee
where $p$ is any point in $B_R$ which does not belong to ${\cal S}$.
This combined with the maximum principle shows
that (\ref{eq:sumpoles})
are the most general harmonic functions leading to regular metrics.
It also follows from (\ref{definitions_UU}) and the positivity
of the spinor inner product that $a_m$ and $\tilde{a}_n$ in
(\ref{eq:sumpoles}) are all non negative.

The spinors $\alpha_A, \beta_{A'}$ and their conjugates give a preferred
basis for the space $\Lambda^2(M)$ of two forms.
The anti-self dual two forms are given in terms of $\alpha_A$ by
\be
\mbox{Re}(\alpha_A\alpha_B\epsilon_{A'B'}), \qquad
\mbox{Im}(\alpha_A\alpha_B\epsilon_{A'B'}), \qquad
i\alpha_{(A}\hat{\alpha}_{B)}\epsilon_{A'B'},
\ee
and the self dual two forms are given in terms of $\beta_{A'}$ by analogous
expressions. The three two forms (\ref{OmegaI}) can be expressed in this
basis as
\begin{eqnarray}
\Omega^1+i\Omega^2&=&
-(\alpha_A\alpha_B\epsilon_{A'B'}+ \beta_{A'}\beta_{B'}\epsilon_{AB})\;
 e^{AA'}\wedge e^{BB'},\nonumber\\
\Omega^3&=&
i(\beta_{(A'}\hat{\beta}_{B')}\epsilon_{AB}
-\alpha_{(A}\hat{\alpha}_{B)}\epsilon_{A'B'})\; e^{AA'}\wedge e^{BB'}.
\end{eqnarray}
The spinor expressions for (\ref{eq:domega}) can now be easily derived using
(\ref{spinor_cond}).

\subsection{Action of the instantons}

The contribution of instantons to physical processes is of course
weighted by their actions. Therefore it is important to evaluate
the actions of the spacetimes we are considering. Previous
computations on this subject should be approached with caution:
there are computational errors in
\cite{Whitt:1984wk} leading to unphysical results such as an action
unbounded from below, while in \cite{Yuille:1987vw} the Maxwell
contribution to the action is not considered. Both of these papers
also work with imaginary electric fields which leads to some
undesirable properties of the actions.

The Riemannian Einstein-Maxwell action, including the Gibbons-Hawking
boundary term, is
\be\label{eq:plainaction}
S = - \int_{M_4} d^4x \sqrt{g^{(4)}} \left[ R^{(4)} - F_{a b} F^{a b} \right] - 2
\int_{\pa M_4}
d^3x \sqrt{\gamma} {\cal K} \,,
\ee
where $\gamma$ is the induced metric on the boundary and ${\cal K}$ is the
trace of the extrinsic curvature of the boundary.

Evaluated on the Einstein-Maxwell instantons we are considering, one finds
\be\label{eq:evalaction}
S = - 2 \pi \lim_{r \to \infty} \int_{S^2} d\Omega^2 r^2 \left[
  \frac{(U + \Ub)^2 \pa_r (U \Ub)}{(U \Ub)^2} + \frac{8}{r} \right] \,.
\ee
Here we have introduced spherical polar coordinates $d{\bf x}^2 =
dr^2 + r^2 d\Omega^2$. The expression (\ref{eq:evalaction}) is
divergent and needs to be regularised by substracting off the action
of a reference geometry. This must be done separately for the
Asymptotically Locally
Flat, Euclidean and Robinson-Bertotti cases. We have assumed in
(\ref{eq:evalaction}) that $\t$ is identified with period $4\pi$.

The easiest case is Asymptotic Local Flatness, with $\b = \bb \neq 0$. Here the
background has simply $U = \Ub = \frac{4\pi}{\b}$, giving flat $S^1 \times
\bR^3$ and a vanishing Maxwell field. One finds
\be
\Delta S_{\text{ALF}} = 8 \pi \beta \left(\sum a_m + \sum \ab_n \right) \,.
\ee
Recall that furthermore $\sum a_m = \sum \ab_n + N - \Nb$ in this
case.

The Asymptotically Locally Robinson-Bertotti case is also straightforward.
Here the background is the Robinson-Bertotti spacetime with $\t$
identified, $AdS_2 / \bZ \times S^2$, supported by magnetic flux, that
is $U = \Ub = \frac{\sum a_m}{r} =  \frac{\sum
  \ab_n}{r}$. The regularised action turns out to vanish
\be
\Delta S_{\text{ALRB}} = 0 \,.
\ee

Now consider the Asymptotically Locally Euclidean case, with
$\frac{4\pi}{\b} = 0$ and $\frac{4\pi}{\bb}=1$. The required
background is Euclidean space with anti self dual Maxwell field, that
is $U = \frac{\sum a_m}{r}$ and $\Ub = 1$. Subtracting this
background regularises the gravitational action, but it does not
remove all the divergences from the Maxwell action. The divergence
of the regularised action tells us that we have not imposed the
correct boundary conditions for the Maxwell field with these
asymptotics.

The standard action (\ref{eq:plainaction}) is appropriate for fixing
the potential at infinity: $\delta A_a = 0$. Different boundary
conditions may be implemented by adding a boundary term to the
action. To obtain a finite action for ALE asymptotics we need to add a
boundary term that entirely cancels the bulk Maxwell action when
evaluated on solutions. The required term is
\be\label{eq:maxwellboundary}
\left. S_{\text{ALE}} \right|_{\text{bdy.}} = 2 \int_{\pa M_4} d^3 x
\sqrt{\gamma} A^a F_{ab} n^b \,,
\ee
where $n^b$ is a unit normal vector to the boundary. The resulting
boundary condition is
\be\label{eq:boundarycondition}
A_a \delta (F^{a b} n_b) = \delta A_a F^{a b} n_b
\quad \text{on} \quad \pa M_4 \,.
\ee
Physically this equation corresponds to keeping a certain linear combination of
the charge and potential fixed at infinity.

With the boundary term (\ref{eq:maxwellboundary}) added, the action is
found to be given by
\be
\Delta S_{\text{ALE}} = 16 \pi^2 \sum \ab_n \,.
\ee
At this moment, we do not have a physical understanding of why the
ALE instantons only contribute to processes in which the
particular boundary condition (\ref{eq:boundarycondition}) is
imposed.

\section{Instanton moduli space metric}

The analysis done in section (\ref{killing_sp}) has demonstrated
that the Einstein-Maxwell gravitational instantons with a Killing
spinor have $3(N+\Nb)$ free parameters or moduli. The Euclidean
group in three dimensions can be used to fix six of these, except
in the case when $N + \Nb = 2$, in which case it only fixes five,
due to the axisymmetry. To obtain the moduli space one should also
quotient by the symmetric group $S_N \times S_\Nb$ acting on the
centres. Note that fixing the action then adds a further
constraint on the centres in the Asymptotically Locally Flat and
Euclidean cases.

While computation of the measure and metric on the moduli space of
Yang-Mills instantons is by now a highly developed field, the case
of gravitational instantons in four dimensions appears to have
been less systematically treated in the literature. In two
dimensions of course the measure plays a fundamental role in
string theory. Reflecting this state of affairs, we now give a
fairly general exposition of the formalism needed to compute
moduli space metrics for gravitational instantons in pure gravity
and Einstein-Maxwell theory.

\subsection{Inner products}

Let us recall the Yang-Mills procedure, but work with just the
$U(1)$ Maxwell case both for simplicity and because this is what we
need anyhow. One begins by writing down a natural ultralocal inner product
on the space of field perturbations. Strictly speaking it is an inner
product on the tangent bundle to the space of fields
\be\label{eq:maxwellinner}
\langle \d A , \d A' \rangle = 2 \int_{M_4} d^4x \sqrt{g} g^{\mu \nu} \d
A_{\mu} \d A'_{\nu} \,.
\ee
In this section it is appropriate to work with spacetime indices
$\mu,\nu\ldots$. One now restricts to considering only
perturbations that are orthogonal to pure gauge transformations.
Thus one requires
\be
0 = \langle \d A , d \Omega \rangle = - 2 \int_{M_4} d^4x \sqrt{g}
g^{\mu \nu} \Omega \nabla_\mu \d A_\nu \,,
\ee
for all $\Omega$. Therefore, perturbations must be considered in
Lorenz gauge
\be\label{eq:lorentz}
\nabla_\mu \d A^{\mu} = 0 \,.
\ee
Given this gauge, we can note that the inner product
(\ref{eq:maxwellinner}) should be thought of as coming from the
quadratic terms in the action. In particular, this
determines the normalisation. The quadratic action is
\bea\label{eq:maxwellact}
S^{(2)}_{\d A} & = & 2 \int_{M_4} d^4x \sqrt{g} \left(
\nabla^\mu \d A^{\rho} \nabla_\mu
\d A_{\rho} - \nabla^\mu \d A^{\rho} \nabla_\rho \d A_{\mu} \right) \nonumber \\
 & \rightarrow & - 2 \int_{M_4} d^4x \sqrt{g} g^{\rho
\sigma} \d A_{\rho} \nabla^2 \d A_{\sigma} + \text{non-derivative terms}\,.
\eea
Where the arrow denotes imposition of the Lorenz gauge. We can see
that the index structure of the gauge field is now that of the
inner product (\ref{eq:maxwellinner}). That is to say, the term in
the last line of (\ref{eq:maxwellact}) is just $-\langle
\d A, \nabla^2 \d A \rangle$, where $\nabla^2$ should be regarded as
an operator on $M_4$.  In this way the inner
product is inherited from the action. The metric on the moduli
space is obtained by restricting the inner product
(\ref{eq:maxwellinner}) to zero modes. To summarise the logic: the
metric on the moduli space is inherited from the quadratic kinetic
terms in the action written in a specific gauge. However, that
gauge must simultaneously imply that field fluctuations are
orthogonal to pure gauge transformations.

We should note at this point that imposing orthogonality to gauge
transformations, with a consequent choice of gauge imposed, is not
completely essential. However, it does greatly simplify instanton
computations and gives a clear physical meaning to the moduli
space metric itself.

For the case of metric fluctuations, there is not a unique ultralocal
inner product with the correct symmetries. Rather we have the
family of de Witt metrics parametrised by $\lambda\in\bR$
\be
\langle \d g , \d g' \rangle_\lambda = \int_{M_4} d^4x G^{\mu \nu \rho
  \sigma}_{(\lambda)} \d g_{\mu \nu} \d g'_{\rho \sigma} \,,
\ee
where
\be\label{eq:dewitt}
G^{\mu \nu \rho \sigma}_{(\lambda)} = \frac{1}{8} \sqrt{g}
\left[g^{\mu \rho} g^{\nu
\sigma} + g^{\mu \sigma} g^{\nu \rho} - 2 \lambda g^{\mu \nu} g^{\rho \sigma}
\right] \,.
\ee
Thus $\lambda$ parametrises the possible inner products. The
metric is positive definite for $\lambda<1/4$ and non-degenerate
for $\lambda\neq 1/4$. In Appendix D we demonstrate that different
values of $\lambda$ indeed give non-equivalent inner products on
moduli space.

The de Witt metric with $\lambda=1$
also appears in Hamiltonian treatments of gravity.
This is not what we are doing here; the metric we want is on four
dimensional Riemannian geometries. In the case of pure gravity
there is a connection, as the four dimensional Euclidean theory
can be lifted to $4+1$ Einstein theory. The gravitational
instantons become Kaluza-Klein monopoles in five dimensions. In
this context the moduli space on the multicentred Gibbons-Hawking
spaces has been computed as the slow motion moduli space metric of
the Kaluza-Klein monopoles \cite{Ruback:1986ag}. We will describe
a lift of our solutions in a later section, but for the moment we
are pursuing a four dimensional treatment.

The ambiguity in the inner product translates into a choice of
gauge. Imposing orthogonality to pure gauge transformations now
requires
\be
0 = \langle \d g , {\mathcal{L}}_{\xi} g \rangle_\lambda =
- \frac{1}{2} \int_{M_4} d^4x \sqrt{g}
\xi^{\mu} \left[\nabla^{\nu} \d g_{\mu \nu} - \lambda \nabla_{\mu} \d
  g^{\nu}{}_{\nu} \right] \,.
\ee
Here ${\mathcal{L}}$ is the Lie derivative. Therefore, metric
fluctuations must be considered in the gauge
\be\label{eq:metgauge}
\nabla^{\nu} \d g_{\mu \nu} = \lambda \nabla_{\mu} \d
  g^{\nu}{}_{\nu} \,.
\ee
In Appendix D we discuss the extent to which the different choices
of $\lambda$ lead to isometric inner products. The result will
certainly not depend on $\lambda$ if all fluctuations are trace
free. All the gauges are equivalent in that case. Indeed, for
noncompact gravitational instantons, all normalisable zero modes
are trace free. This is not true for the compact gravitational
instanton, K3. However, we now need to check compatibility with
the quadratic kinetic terms in the action. The quadratic action,
only keeping track of derivative terms, is
\bea\label{eq:quadraticgravity}
S^{2}_{\d g} & = & \frac{1}{4} \int_{M_4} d^4x \sqrt{g} \left(
\nabla^\mu \d g^{\rho \sigma} \nabla_\mu \d g_{\rho
\sigma} - \nabla^\mu \d g^\rho{}_\rho \nabla_\mu \d
g^\sigma{}_\sigma \right. \nonumber \\
 & & \qquad \qquad \qquad \left. - 2 \nabla^\mu \d g^\rho{}_\mu \nabla^{\sigma} \d g_{\rho \sigma}
+ 2 \nabla^\mu \d g^\rho{}_\rho \nabla^{\sigma} \d g_{\mu \sigma} \right)\\
 & \rightarrow & -\frac{1}{8} \int_{M_4} d^4x \sqrt{g}
\left[g^{\mu \rho} g^{\nu
\sigma} + g^{\mu \sigma} g^{\nu \rho} - 2 (1+2\lambda^2 - 2 \lambda) g^{\mu \nu} g^{\rho \sigma}
\right] \delta g_{\mu \nu} \nabla^2 \delta g_{\rho
\sigma}\,\nonumber,
\eea
where arrow denotes imposition of the gauge (\ref{eq:metgauge}).
Generically, this does not correspond to the de Witt
(\ref{eq:dewitt}) inner product which we started with. For
consistency, we now need to impose $2 \lambda^2 - 3 \lambda + 1
=0$. The two solutions to this equation are $\lambda = 1$ and
$\lambda = \frac{1}{2}$. These are in fact rather interesting
values. The first is that obtained from viewing the instanton
moduli space as the slow motion moduli space of 4+1 dimensional
Kaluza-Klein monopoles \cite{Ruback:1986ag}. The second
corresponds to de Donder gauge, perhaps the most natural gauge for
the theory, and was considered recently because gradient flow on
the space of metrics with this inner product is Ricci flow
\cite{Headrick:2006ti}.

It follows from the previous few paragraphs that for gravitational
instantons there are two preferred gauges, which correspond to
taking $\lambda = 1$ or $\lambda =
\frac{1}{2}$ in the de Witt metric. However, we are interested in
Einstein-Maxwell theory, so we furthermore need to take into
account the fact that the Maxwell field also transforms under
infinitessimal diffeomorphisms $A \to A + {\mathcal{L}}_{\xi} A$.
Orthogonality to such diffeomorphisms therefore requires
\be\label{eq:bothgauge}
\langle \d g , {\mathcal{L}}_{\xi} g \rangle_\lambda + \langle \d A ,
    {\mathcal{L}}_{\xi} A \rangle = 0 \,.
\ee
Using the Lorenz condition on the gauge field perturbation
(\ref{eq:lorentz}) one finds that the orthogonality condition
(\ref{eq:bothgauge}) requires that the following gauge be
implemented for the moduli
\be\label{eq:gandA}
\nabla^{\nu} \d g_{\mu \nu} - \lambda \nabla_{\mu} \d
  g^{\nu}{}_{\nu} = - 4 \d A^\nu F_{\nu \mu} \,.
\ee
Once again, we need to substitute this gauge choice into the
quadratic term of the action. This is similar to the case
of pure gravity (\ref{eq:quadraticgravity}) except that there are
two extra terms due to the right hand side of the gauge condition
(\ref{eq:gandA}). One of these does not involve any derivatives of
$\d A_\mu$ or $\d g_{\mu \nu}$ and so does not contribute to the
quadratic terms. However, the other term involves a single
derivative. This latter term is always present unless $\lambda =
\frac{1}{2}$, suggesting that this is the preferred gauge for
Einstein-Maxwell instantons.


\subsection{Towards the moduli space metric}

To find the moduli space metric we need to find the general
solution to the linearised Einstein-Maxwell equations satisfying
the gauge conditions (\ref{eq:lorentz}) and (\ref{eq:gandA}). Once
we have the solution, we should then evaluate the norm of the
fluctuations using the results of the previous section. Given that
we have the general solution at a nonlinear level, we can easily
solve the linearised Einstein-Maxwell system by perturbing the
full solutions. However, these solutions will not be in the
required gauge. Finding a gauge transformation to map the solution
into the correct gauge does not appear easy.

An alternative and more elegant approach is that employed in
\cite{Ruback:1986ag} to find the moduli space metric on the
Gibbons-Hawking gravitational instantons. This uses the existence
of $N$ closed self dual two forms on the background, $F^J$, as
well as the three self dual K\"ahler forms $\Omega^i$ to
write the metric fluctuation
\be
\d g^{iJ}_{\mu\nu} = \Omega^{i\rho}{}_{(\mu} F^J_{\nu) \rho} \,.
\ee
This perturbation solves the linearised Einstein
equations. Furthermore, it is transverse and tracefree and therefore
solves the gauge condition required for pure Einstein gravity.

Note that this approach combines supersymmetry,
which provides the three K\"ahler forms, and the topology of solution,
which has $b^+_2 = N$ and hence implies the existence of the closed self dual
forms $F^J$. Using these modes, \cite{Ruback:1986ag} shows that the
moduli space metric is given in terms of intersection matrix of the
Gibbons-Hawking background and is flat.

So far, we have not been able to adapt this argument to the
Einstein-Maxwell case in a way consistent with the gauge condition
(\ref{eq:gandA}). We hope that the framework presented in this
section will be a useful starting point for future work on the
moduli space metric.

\section{Lift to five dimensions}

\subsection{Lifting the solutions}

Recall the following feature of field theory instantons:
instantons in $D$ dimensions may be viewed as solitons in $(D+1)$
dimensions. Furthermore, the $L^2$ instanton metric coincides with
a natural Riemannian metric on the moduli space of solitons that
is induced from the kinetic term in the $(D+1)$ dimensional
action. This is interesting given the differing interpretations of
the metrics in each case. The metric is relevant at the classical
level in $(D+1)$ dimensions, as its geodesic motion approximates
the soliton dynamics in the nonrelativistic limit
\cite{Manton:1981mp}. However, in $D$ dimensions the metric is only
important in quantum field theory, where measures on solution
spaces are needed.

This procedure can also be applied to the 4 dimensional
Einstein-Maxwell gravitational instantons (\ref{IWmetric}) if it
is possible to lift them to Lorentzian metrics which are solitons
of some theory in higher dimensions. Of course the resulting
moduli space metric could depend on the choice of higher
dimensional theory. In this section we study one possible theory
in $(4+1)$ dimensions. The five dimensional metrics resulting from
the lift are interesting in their own right, and we clarify some
of their properties in this section. In the following section
\ref{Section_metric} we shall discuss the metric on the slow motion
moduli space of these solitons.

Einstein-Maxwell theory without a dilaton cannot be consistently
lifted to pure gravity in five dimensions\footnote{The need
for a dynamical scalar field was not originally
appreciated in the 1920s by Kaluza and Klein who set it to a constant.
This mistake was corrected more than 20 years latter by Jordan and Thiry.}.
However, Einstein-Maxwell
configurations may be lifted to solutions of five dimensional
Einstein-Maxwell theory with a Chern-Simons term. This lift is the
bosonic sector of the lift from ${\mathcal{N}}=2$ supergravity in
four dimensions to ${\mathcal{N}}=2$ supergravity in five
dimensions \cite{Chamseddine:1980sp,Lozano-Tellechea:2002pn}.
We are interested in lifting the four dimensional Riemannian theory to
a Lorentzian theory on a five dimensional manifold $M_5$.
The four dimensional action is
\be
S_4 = \int d^4x \sqrt{g^{(4)}} \left[ R^{(4)} - F_{a b} F^{a b} \right] \,,
\ee
with equations of motion given by (\ref{eq:4deqns}). The five
dimensional action is
\be
\label{SEMCS}
S_5 = \int d^5x \sqrt{-g^{(5)}} \left[R^{(5)} - H_{\a \b} H^{\a \b} \right]
- \frac{8}{3\sqrt{3}} \int H \wedge H \wedge W \,,
\ee
where $H = dW$ is the five dimensional Maxwell field. We use greek
indices ranging from $0$ to $4$ in five dimensions. The equations of
motion in five dimensions are
\bea\label{eq:5deqns}
G_{\a \b} & = &  2 H_{\a}{}^{\g} H_{\b \g} - \frac{1}{2} g^{(5)}_{\a \b} H^{\g
  \d} H_{\g \d} \,, \nonumber \\
d \star_5 H & = & - \frac{2}{\sqrt{3}} H \wedge H \,.
\eea

Given a solution, $g^{(4)}$ and $F=dA$, to the four dimensional
equations (\ref{eq:4deqns}),  we may lift the solution to five dimensions as follows:
\bea\label{eq:5dconfig}
g^{(5)} & = & g^{(4)} - (dt + \Phi)^2 \,, \nonumber \\
W & = & \frac{\sqrt{3}}{2} A \,,
\eea
where $\Phi$ is a one form determined by $g^{(4)}$ and $F$ through
\be\label{eq:phieqn}
d \Phi = \star_4 F \,.
\ee
One may then check that the five dimensional configuration
(\ref{eq:5dconfig}) solves the equations of motion
(\ref{eq:5deqns}). Note that solutions to (\ref{eq:phieqn}) exist
because $d \star_4 F = 0$ on shell. In our case we may solve for
$\Phi$ explicitly to find
\be
\Phi = - \frac{1}{2} \left(U^{-1} + \Ub^{-1}\right) (d\t + {\bf \w}) + {\bf \chi} \,,
\ee
where $\chi$ satisfies
\be
\nabla \times {\bf \chi} = \frac{1}{2} \nabla \left(U - \Ub \right) \,.
\ee

The supersymmetric solutions of ${\mathcal{N}}=2$ supergravity in five
dimensions have been classified \cite{Gauntlett:2002nw}. For the case
of a timelike Killing spinor the general solution is given as a $U(1)$
fibration over a four real dimensional hyperK\"ahler manifold.
It was shown in \cite{Gauntlett:2002nw} how the lift of the Lorentzian
Israel-Wilson-Perj\'es solutions to five dimensions could be expressed as a
fibration over the multicentred Gibbons-Hawking metrics \cite{Gibbons:1979zt}.

It turns out that the lift of the Riemannian
Israel-Wilson-Perj\'es solutions we are considering may also be
expressed as a fibration over the multicentred Gibbons-Hawking
metrics. The five dimensional metric (\ref{eq:5dconfig}) can be
written as follows\footnote{Writing the spacetime in the form
(\ref{eq:goodform}) locates the five dimensional solution in the
classification of \cite{Gauntlett:2002nw}. In section 3.7 of that
paper the general supersymmetric fibration over a Gibbons-Hawking
base with $\pa/\pa t$ a Killing vector is given in terms of three
harmonic functions. For our solution these correspond to $L = 2
\Ub$, $K = - \Ub$ and $M = - 2 \Ub$.}
\be\label{eq:goodform}
g^{(5)} = - f^2 \left(d\t + \w' \right)^2 + f^{-1} g^{GH} \,,
\ee
where the Gibbons-Hawking metric is
\be
g^{GH} = V^{-1} \left(dt + \chi \right)^2 + V d{\bf x}^2 \,,
\ee
with harmonic funtion
\be
V = \frac{1}{2} \left(U - \Ub \right)\,.
\ee
The remaining functions in the metric (\ref{eq:goodform}) are
\be
f = \frac{V}{U \Ub} \,,
\ee
and
\be
\w' = {\bf \w} - \frac{1}{2 f^2} \left(U^{-1} + \Ub^{-1} \right)
\left(dt + {\bf \chi} \right) \,.
\ee
Note that the hyperK\"ahler base itself is in general not regular,
even changing signature at points where $U = \Ub$. This is
perfectly compatible with regularity of the five dimensional
spacetime.

The case $U = \Ub$ is exceptional and cannot be written in the
form (\ref{eq:goodform}). Instead, these metrics have null
supersymmetry in five dimensions. The metric is\footnote{The
metric (\ref{eq:nullform}) falls within the classification of
\cite{Gauntlett:2002nw} for spacetimes with null supersymmetry by
setting their functions $H = - {\mathcal{F}}=U$ and
${\mathbf{a}}=0$.}
\be\label{eq:nullform}
g^{(5)} = \frac{2 dt d\t}{U} - dt^2 + U^2 d{\bf x}^2 \,.
\ee

\subsection{Regularity and causality}

The interesting points in the five dimensional metric are the
centres where $U \to \infty$ or $\Ub \to \infty$. In the four
dimensional Riemannian Israel-Wilson-Perj\'es solutions these can
always be made to be regular points \cite{Whitt:1984wk,
Yuille:1987vw} as we reviewed above. We need to re-examine the
regularity of the metric around these points and also check for the
possible occurrence of closed timelike curves.

Before zooming in on the centres note the following. Firstly, that
\be
{g^{(5)}}_{\t\t} \equiv g^{(5)}\Big(\frac{\pa}{\pa \t}, \frac{\pa}{\pa \t}\Big)
= - \frac{(U-\Ub)^2}{(2 U \Ub)^2} < 0 \,,
\ee
if $U \neq \Ub$. Therefore, to avoid closed timelike curves throughout
the five dimensional spacetime we must not identify $\t$.
Secondly, possible candidates for the location of horizons are where
the metric becomes degenerate
\be\label{eq:horizons}
0 = {g^{(5)}}_{tt} {g^{(5)}}_{\t\t} - [{g^{(5)}}_{t\t}]^2 = -\frac{1}{U \Ub} \,.
\ee
This occurs at the centres where $U$ or $\Ub$ diverge.

In order to understand the geometry near the centres,
there are three different cases we need to consider separately.
The first is that $U \to \infty$ while $\Ub$ remains finite. Using
polar coordinates $(r=\rho^2/4, \theta, \phi)$  centred on the point
${\bf x}_m$ and
requiring that $a_m \Ub({\bf x}_m) = 1$, the metric becomes
\be
ds^2 = d\rho^2 + \frac{\rho^2}{4} \left[(d\t + \cos\theta d\phi)^2
+ d\Omega^2_{S^2} \right] - \left(dt - a_m d\t/2
\right)^2
\ee
as $\rho \to 0$, with $d\Omega^2_{S^2}=d\theta^2+\sin^2\theta d\phi^2$.
The metric
may be made regular about this point if we identify $\t$ with period
$4\pi$. Unfortunately this introduces closed timelike curves as we
discussed. If we choose not to identify $\t$ we are left with
timelike naked singularities at the centres. We see that there is no
horizon at these points, but rather a (singular) origin of polar
coordinates. Therefore, metrics with this behaviour at the centres
cannot lift to causal, regular solitons in five dimensions.

The remaining two possibilities involve coincident centres where
both $U$ and $\Ub$ go to infinity, so that ${\bf x}_m = {\bf \xb}_m$. One
needs to treat separately the cases where $a_m = \ab_m$ and where
$a_m \neq \ab_m$. In the latter case we again find regularity at
the expense of closed timelike curves going out to infinity, or
alternatively naked singularities. This
leaves only the former case with $a_m = \ab_m$ for all $m$. That is,
$\Ub = U + k$, with $k$ some constant.

By considering the asymptotic regime, one can see that in order to
obtain a sensible asymptotic geometry without closed timelike
curves, one requires that either both $U$ and $\Ub$ go to a
constant at infinity or they both go to zero. Rescaling the
harmonic functions and performing a duality rotation on the
Maxwell field, as we discussed in four dimensions above, implies
that without loss of generality $U=\Ub$. We consider this case in
the following subsection.

\subsection{Multi solitonic strings}

The only lift that leads to a globally regular and
causal five dimensional spacetime is the case $U=\Ub$, which corresponds to
the Euclidean Majumdar-Papapetrou metric in four dimensions.
The metric is (\ref{eq:nullform}), with a null Killing spinor.
Away from the centres, the spacetimes approach either
$\bR^{1,4}$ or $AdS_3 \times S^2$, with $U$ going to a constant or
zero at infinity, respectively.

With a rescaling of coordinates, the geometry near the centres
where $U \to \infty$ may be written
\be\label{eq:ads}
ds^2 = a_m^2 \left[\frac{dr^2}{r^2} + 2 r dt d\t - dt^2 +
d\Omega^2_{S^2} \right] \,.
\ee
Calculating the curvature shows that
this metric locally describes $AdS_3 \times S^2$. One might be tempted
to conclude that this represents the near horizon
geometry of an extremal black string in five dimensions. However, the
coordinates (\ref{eq:ads}) are a little unusual, the sign of
$dt^2$ differing from the metric of an extremal BTZ black hole
\cite{Banados:1992gq}.  In particular, the Killing vector $\pa/\pa t$
is everywhere regular and timelike. This remains true in the full
spacetime (\ref{eq:nullform}). There is no horizon and the
degeneration of the metric at the centres is analogous to an origin of
polar coordinates.

The coordinates in (\ref{eq:ads}) may be mapped to Poincar\'e coordinates as follows
\bea\label{eq:poincare}
Y & = & \frac{1}{r^{1/2}\cos \frac{t}{2}} \,,\nonumber \\
X & = & \frac{\t}{2} - \frac{1}{2} \left[\frac{1}{r}
  - 1 \right] \tan \frac{t}{2} \,,
\nonumber \\
T & = & \frac{\t}{2} - \frac{1}{2} \left[\frac{1}{r} +
1 \right] \tan \frac{t}{2} \,,
\eea
so that the metric becomes
\be
ds^2 = \frac{4 a_m^2}{Y^2} \left(-dT^2 + dX^2 + dY^2 \right) + a_m^2
d\Omega^2_{S^2} \,.
\ee
There is no singularity at $t = \pm \pi$ as may be checked by
writing down the embedding of $AdS_3$ as a quadric in $\bR^{2,2}$
in terms of these coordinates. The map (\ref{eq:poincare}) is
periodic in $t$. Taking $t$ with infinite range corresponds to
passing to the (causal) universal cover of $AdS_3$. There is no
need to identify $\t$ and therefore the spacetime is causal.

The metrics (\ref{eq:nullform}) give causal, regular
solutions to the five dimensional theory with an everywhere defined
timelike Killing vector. Writing the metric in the form
\be
g^{(5)} = -(dt - d\t/U)^2 + \frac{d\t^2}{U^2} + U^2 d{\bf x}^2
\,,
\ee
suggests that the spacetimes should be thought of as containing
$N$ parallel `solitonic strings'. The strings have worldvolumes in
the $t-\tau$ plane. There is a plane fronted wave
\cite{Gauntlett:2002nw} carrying momentum along the $\pa/\pa \t$
direction of the string. We call these plane fronted waves
solitonic strings to emphasise that the fields are localised along
strings and there are no horizons. The strings are magnetic
sources for the two form field strength
\be
H = - \sqrt{3} \star_3 d U \,.
\ee
This is possible because of the topologically nontrivial $S^2$ at
each centre (\ref{eq:ads}).

We end this subsection by remarking that any solution to
Einstein-Maxwell-Chern-Simons theory (\ref{eq:5deqns}) in $4+1$
dimensions can be lifted to a solution to 11 dimensional
supergravity given by the product metric of $g^{(5)}$ and a flat
metric on the six torus. The eleven dimensional four form is given
by $H\wedge\Omega_T$, where $\Omega_T$ is the K\"ahler form on the
torus. We have not pursued here an M theory interpretation of
these solutions.

\section{Slow motion in five dimensions}
\label{Section_metric}

An interesting feature of BPS solitons is the cancelation between
forces which makes static multi-soliton configurations possible.
This is clear for the 3+1 dimensional
Majumdar-Papapetrou multi black holes, where the electrostatic repulsion
is balanced by gravitational attraction. These black holes are in this sense
analogous to a nonrelativistic system of massive charged particles,
with the charge-to-mass ratio chosen to balance the Newtonian attraction and
Coulomb repulsion.

The nature of the forces in the, stationary but not static, 4+1
dimensional solution (\ref{eq:5dconfig}) is presumably more
complicated. We shall not study this problem here, and instead
focus on the scattering of slowly moving solitons. The question we
are interested in is whether there is a direct connection between
the metric on the moduli space of four dimensional instantons and
the metric on the moduli space describing slow motion of the 4+1
dimensional solitons. The metrics do coincide for pure gravity
instantons \cite{Ruback:1986ag}.

One can follow Manton's method for truncating the infinite number
of degrees of freedom of the gravitational field to the finite
dimensional moduli space ${\cal M}$ of solitons\footnote{In this
section we will refer to any of the solutions (\ref{eq:5dconfig})
as solitons, even if they are singular or contain closed timelike
curves. Part of our motivation is to compare with the moduli space
metric of four dimensional gravitational instantons
(\ref{IWmetric}) where everything is regular, even if $U \neq
\Ub$.}. This means that we shall be neglecting both gravitational
and electromagnetic radiation, and consider only velocity
dependent forces which perturbed solitons induce on each other. As
for the four dimensional instantons, the space ${\cal M}$ is not
the whole of $\bR^{3(N+{\tilde N})}$. To obtain ${\cal M}$ we need
to quotient by the permutation group $S_N\times S_{\tilde N}$, and
the Euclidean group in three dimensions.

By considering the slow motion approximation to the initial value
formulation of 4+1 dimensional Einstein-Maxwell-Chern-Simons
(EMCS) theory, one can find the moduli space metric from the
effective action where the field degrees of freedom have been
integrated out. In the moduli space approximation the centres
become functions of $t$ and geodesic curves $\{{\bf x}_m(t), {\bf
\tilde{x}}_n(t)\}$ correspond to slow motion of a multi solitonic
string configuration.

The initial data for EMCS theory (\ref{SEMCS}) consists of a four
dimensional manifold $\Sigma$ together with a Riemannian metric
$\gamma_{\mu\nu}$, a symmetric tensor $K_{\mu\nu}$, a two form $B$
and a one form $E$. Given a metric $g^{(5)}$ and a one form
potential $W$ on $M_5$ we can perform a $4+1$ decomposition if
there exist a function $t$ whose gradient is everywhere timelike.
In this case $\Sigma$ is a level set of $t$, and we choose adapted
local coordinates $(t, x^a)$ such that the normal to $\Sigma$
takes the form
\be
{\cal N}=N^{-1}(\pa_t-N^\mu\pa_\mu)\,,
\ee
where $N$ and $N^\mu$ are the lapse function and the shift vector.
The spatial metric $\gamma_{\mu\nu}$ and the second fundamental
form $K_{\mu\nu}$ can now be read off from the formulae
\begin{eqnarray}\label{eq:ans1}
g^{(5)}&=&-N^2 dt^2+\gamma_{\mu\nu}(dx^\mu+N^\mu dt)(dx^\nu+N^\nu dt)\,, \nonumber\\
K_{\mu\nu}&=&\frac{1}{2}N^{-1}(\pa_t \gamma_{\mu\nu}-D_\mu
N_\nu-D_\nu N_\mu)\,,
\end{eqnarray}
where $D$ is the covariant derivative compatible with $\gamma$ on $\Sigma$.
We also decompose the one form $W$ and two form $H=dW$ as
\be\label{eq:ans2}
W= W_0 N dt + W_\mu dx^\mu \,, \qquad H=E \wedge N dt + B\,.
\ee
This last formula implies expressions for $E$ and $B$ as exterior
derivatives of the potentials $W_0$ and $W_\mu$.

The next step is to implement the 4+1 decomposition at the level of
the action. After neglecting a total derivative term, the following
action is obtained from substituting (\ref{eq:ans1}) and
(\ref{eq:ans2}) into the EMCS action (\ref{SEMCS})
\bea\label{eq:decomaction}
S_{4+1} & = & \int d^4x dt N \sqrt{\gamma}
\left[R^{\gamma}+K_{\mu\nu}K^{\mu\nu}-K^2\right] \nonumber \\
 & + & \int d^4x dt N \sqrt{\gamma} \left[2 E_\mu E^\mu - B_{\mu\nu} B^{\mu\nu} +
  2 B_{\mu\nu} B_\rho{}^{\nu} \frac{N^\mu N^\rho}{N^2} \right] \nonumber \\
 & - & \frac{8}{3\sqrt{3}} \int \left[W_0 B \wedge B - 2 B \wedge E
  \wedge W_\mu dx^\mu \right] \wedge N dt \,.
\eea
Here $R^{\gamma}$ is the Ricci scalar of $\gamma$, $K=\gamma^{\mu
\nu}K_{\mu\nu}$, and all contractions use the metric $\gamma$. The three
lines come from the Einstein-Hilbert, Maxwell and Chern-Simons
terms in the action (\ref{SEMCS}), respectively. If we think of
the expression (\ref{eq:decomaction}) as an action for the fields
$\{\gamma_{\mu\nu},W_\mu,W_0,N_\mu,N\}$, then we see that the last
three of these appear without time derivatives. They are Lagrange
multipliers and impose the constraints of conservation of energy,
momentum and charge
\be\label{eq:constraintx}
\frac{\d S_{4+1}}{\d N} = \frac{\d S_{4+1}}{\d N_\mu} = \frac{\d
  S_{4+1}}{\d N W_0} =
0 \,.
\ee
Arbitrary initial data will not evolve to a solution of the EMCS
theory. One needs to impose the constraint equations (\ref{eq:constraintx}).

To consider the slow motion dynamics of a perturbed stationary
solution, we allow the moduli to become time dependent and work to
first order in the velocities
\be
v^J = \frac{d x^J}{dt},
\ee
where we have used $x^J$ to denote a general modulus. This induces
a time dependence in the solution which to first order can be
written
\be\label{eq:timederiv}
\frac{d \g_{\mu\nu}}{dt} = \d \g^J_{\mu\nu} v^J \,,
\qquad \frac{d W_{\mu}}{dt} = \frac{\sqrt{3}}{2} \d A^J_{\mu} v^J \,,
\ee
where $\d \g^J_{\mu\nu}, \d A^J_{\mu}$ is the zero mode
corresponding to the modulus $x^J$. In general, simply allowing
the moduli to depend on time will not give a spacetime that solves
the constraint equations, even to first order in the velocities.
Instead, it will be necessary to add extra terms linear in the
velocities to the original solution. An early example of this
technique in gravity is the slow motion of Majumdar-Papepetrou
black holes \cite{Ferrell:1987gf}.

For the case of the Kaluza-Klein monopole lift of the
Gibbons-Hawking solutions, it turns out that it is sufficient to
simply promote the moduli to time dependent fields. The constraint
equations are automatically solved to first order in velocities
\cite{Ruback:1986ag}. This lies behind the simple identification
of the moduli space metrics in four and five dimensions. Let us
see whether the constraint equations are solved in our case.

To first order in velocities, the charge conservation and momentum
conservation constraints become
\be\label{eq:conservation}
D^\mu (\d A_\mu^J/N) = 0 \,, \qquad D^\mu (\d\g_{\mu\nu}^J/N) =
D_\nu (\d \g^{J \mu}{}_{\mu}/N) \,.
\ee
Here we used (\ref{eq:timederiv}). It is interesting to see that
these two constraints take the form of gauge conditions. They may
be imposed on the moduli fields and no extra terms are necessary.
Although these gauge conditions look similar to those encountered
in section 3.1, they are quite different. The choice of time
slicing is not the same. By comparing (\ref{eq:5dconfig}) and
(\ref{eq:ans1}) we see that $\g_{\mu\nu} = g_{\mu\nu} -
\Phi_{\mu} \Phi_{\nu}$. Working through the changes to the
covariant derivative shows that the charge conservation constraint
in (\ref{eq:conservation}), for instance, becomes
\be
\nabla_\mu \left( \left[ (1-\Phi^2) g^{\mu \nu} + \Phi^\mu \Phi^\nu \right] \d A_\nu^J \right)  = 0\, .
\ee
Deriving this expression uses $1/N^2 = 1 - \Phi^2$. Here $\Phi^2$
is contracted with $g_{\mu \nu}$. A similar expression exists for
the momentum constraint. It is clear that this is not the Lorenz
gauge that we used for the instanton moduli space. As discussed,
the instanton moduli space metric is gauge dependent. This is the
first indication that there is not a direct connection between the
instanton and soliton moduli space metrics for our solutions.

A more significant problem arises from the Hamiltonian constraint.
To first order in velocities the constraint is
\be\label{eq:hamiltonian}
\d g^J_{\mu\nu} D^\mu N^{\nu} - \d g^{J \mu}{}_\mu D^{\nu} N_{\nu}
= \frac{N}{\sqrt{\g}} \e^{\mu \nu \rho \sigma} F_{\mu \nu} \d
A^J_\rho A_{\sigma}\,.
\ee
This is an algebraic relation between the various metric and
Maxwell field moduli. We might hope that (\ref{eq:hamiltonian}) is
solved for all moduli
for $\lambda=1$.
Unfortunately, it is
clear that this will not work. Notice that the Hamiltonian
constraint (\ref{eq:hamiltonian}) involves a symmetric derivative
of $N^\mu$. This translates into a symmetrised derivative of
$\Phi^\mu$. However, only the antisymmetrised derivative of
$\Phi^{\mu}$ can be expressed in terms of the four dimensional
fields via (\ref{eq:phieqn}). The Hamiltonian constraint will
require extra modes to be turned on for a consistent time
dependent solution.

The upshot of this section is therefore that, unlike in case of
Yang-Mills instantons or pure gravitational instantons, the slow
motion moduli space metric of the five dimensional soliton cannot
be directly reduced to the four dimensional instanton moduli space
metric. A full blooded computation of the backreaction of the
moduli velocities onto the spacetime is necessary.

\section{Discussion}

In this paper we have discussed various properties of multi-instanton
solutions of Euclidean Einstein-Maxwell theory. We have also shown how
these solutions may be lifted to `solitonic string' solutions of five dimensional Lorentzian
Einstein-Maxwell-Chern-Simons theory. There are roughly three types of
application for the solutions we have discussed. We hope that the present work
has provided a solid base for future investigations.

Firstly and perhaps most interestingly, given that the instantons only involve fields
that are observed to exist in nature, would be to understand the physical
effects mediated by these solutions. A well known example of the physical effect
of Euclidean Einstein-Maxwell theory is the bounce that describes the pair creation of
charged black holes in a sufficiently strong electromagnetic field \cite{Garfinkle:1990eq}.
One possible direction of study would be to ask whether the instantons tell us anything about the structure of the vacuum of Einstein-Maxwell theory, say as a function of temperature.

Secondly, it would be of interest to understand the role of these solutions as supersymmetric
building blocks within string and M theory. Either as higher dimensional supergravity instantons
\cite{deVroome:2006xu}, or as a component of Lorentzian compactification or brane solutions.
This would be analogous to the ubiquitous appearance of the Gibbons-Hawking metrics
in higher dimensions.

Thirdly, there are various mathematical aspects that we have not developed completely.
Some of these have physical consequences. It is important to understand the index theory
associated with the zero modes of the instantons. This will determine which correlators the instantons contribute to and also their effect on topological terms in the Lagrangian. Furthermore, we have not discussed determinants of quadratic fluctuations about the solutions. An interesting question is whether supersymmetry is sufficient in this case to force the one loop determinants to cancel.

On a slightly different note, a completely distinct set of Einstein-Maxwell instantons may be
constructed. LeBrun has found explict multicentred scalar flat K\"ahler metrics \cite{L91}.
These give solutions to Einstein-Maxwell theory with the field strength given by half the
K\"ahler form plus the Ricci form. It would interesting to study these solutions in more depth
and elucidate their relation, if any, with the solutions that we have discussed.

\section*{Acknowledgements}

We would like to thank David Berman, Roberto Emparan, Gary
Gibbons, Jan Gutowski, Matt Headrick, Hari Kunduri, James Lucietti, David
Mateos, Malcolm Perry, Simon Ross, Paul Tod and David Tong for
helpful comments at various points during this work.

This project was begun while SAH was supported by a research
fellowship from Clare College Cambridge. His research was
supported in part by the National Science Foundation under Grant
No. PHY99-07949.

\appendix

\section{Expressions for the potentials}

An explicit formula for ${\bf \w}$ may be obtained from
integrating (\ref{IWequations}). There is a choice of gauge
involved as ${\bf
\w}$ is only defined up to gradient. We can see that the
contributions to ${\bf \w}$ will come from cross terms in the sums
defining $U$ and $\Ub$ (\ref{eq:sumpoles}). Therefore we can write
\be
{\bf \w} = \sum_{m n} {\bf \w}_{mn} - \frac{4 \pi}{\beta} \sum_n
\tilde
{\bf \w}_n +
\frac{4 \pi}{\tilde \beta} \sum_m {\bf \w}_m \,.
\ee
A possible form for the first term $\w_{mn}$ is
\bea\label{eq:wdouble}
{\bf \w}_{mn} & = & - a_m \ab_n \frac{({\bf x} - {\bf x}_m) \cdot
({\bf x} - {\bf \xb}_n)}{\mid {\bf x} - {\bf x}_m \mid \mid {\bf
x} - {\bf \xb}_n
\mid} \frac{({\bf x}_m - {\bf \xb}_n) \times
({\bf x} - ({\bf x}_m + {\bf \xb}_n)/2)}{\mid ({\bf x}_m - {\bf
\xb}_n)
 \times ({\bf x} - ({\bf x}_m + {\bf \xb}_n)/2)\mid^2} \nonumber
\\
 & = & \frac{a_m \ab_n}{\mid {\bf x}_{mn} \mid}
\frac{{\bf x}_{-m} \cdot {\bf x}_{-n}}{\mid {\bf x}_{-m} \mid \mid {\bf x}_{-n} \mid}
  \nabla \left\{ \tan^{-1} \frac{\left[{\bf x}_{mn} \times \left(
 {\bf x}_{mn} \times  ({\bf x}_{-m} + {\bf x}_{-n}) \right) \right]
 \cdot {\bf k}}{\mid {\bf x}_{mn} \mid \left[ {\bf x}_{mn} \times ({\bf x}_{-m} + {\bf x}_{-n}) \right] \cdot
 {\bf k}} \right\}\,.
\eea
In the second expression ${\bf x}_{mn} = {\bf x}_m - {\bf \xb}_n$,
${\bf x}_{-m} = {\bf x} - {\bf x}_m$, ${\bf x}_{-n} = {\bf x} -
{\bf \xb}_n$ and ${\bf k}$ is an arbitrary constant vector. This
breaking of symmetry is the price we need to pay for expressing
part of the term as a gradient.

A possible expression for the remaining terms, writing ${\bf \w}$
as a form for ease of notation, is
\be\label{eq:wsingle}
{\bf \w}_m = a_m \frac{(z-z_m) \left(-(y-y_m) dx + (x-x_m) dy
\right)}{\mid {\bf x} - {\bf x}_m \mid [(x-x_m)^2+(y-y_m)^2]} \,.
\ee
The $\tilde {\bf \w}_n$ are given by the same expression but with
$a_m \to \ab_n$ and ${\bf x}_m \to {\bf \xb}_n$. As is usual, the
choice of gauge for (\ref{eq:wsingle}) necessarily breaks the
rotational symmetry and has Dirac strings.

Both of the previous two formulae are more naturally given in
polar coordinates. However, the angles would depend on the centres
or pairs of centres in question. If we want coordinates that are
valid for all the ${\bf \w}_{mn}$ and ${\bf \w}_m$ at once then we
need to use Cartesian coordinates.

The gauge for ${\bf \w}$ that we have chosen in (\ref{eq:wdouble})
and (\ref{eq:wsingle}) satisfies $\nabla \cdot {\bf \w} = 0$. In
fact, the expression in curly brackets in (\ref{eq:wdouble}) is a
harmonic function.

We may also integrate the field strength (\ref{eq:fieldstrength})
to obtain an explicit potential. This is defined up to a gradient.
A possible expression is
\be\label{eq:decompos}
A = A_4 (d\t + {\bf \w}) + {\bf A}\,,
\ee
with
\be\label{eq:choose}
A_4 = \frac{U - \Ub}{2 U \Ub} \quad \text{and} \quad {\bf A} = -
\frac{1}{2}\left[ \sum_m {\bf \w}_m + \sum_n \tilde {\bf \w}_n
\right] \,.
\ee
Where the ${\bf \w}_m, \tilde {\bf \w}_n$ are as given in
(\ref{eq:wsingle}). More invariantly, $\nabla \times {\bf A} =
-\frac{1}{2} (U + \Ub)$. Note that with this choice of gauge,
$\nabla \cdot {\bf A} = 0$.

\section{Gamma matrix conventions and spin connection}

We work with a chiral representation of the Euclidean gamma matrices
\be
\G^a =
\left(\begin{array}{cc}
0 & -i \s^a \\
i \tilde \s^a & 0
\end{array}\right) \,,
\ee
where $\s^a = (i,{\bf \tau})$ and $\tilde \s^a = (-i,{\bf \t})$. Here
${\bf\tau}$ are
the Pauli matrices. The gamma matrices satisfy $\{\G^a,\G^b\} = 2
\d^{ab}$. We define
\be
\G^{ab} \equiv \frac{1}{2} \left[\G^a, \G^b \right] =
\left(\begin{array}{cc}
\s^{ab} & 0 \\
0 & \tilde \s^{ab}
\end{array}\right) \,,
\ee
where $\s^{ab} = \frac{1}{2} \left[\s^a \tilde \s^b - \s^b \tilde
\s^a
  \right]$ and $\tilde \s^{ab} = \frac{1}{2} \left[\tilde \s^a \s^b - \tilde
  \s^b \s^a  \right]$. As two forms, $\s^{ab}$ is anti-self dual
whilst $\tilde \s^{ab}$ is self dual. Finally, let $\G^5 = \G^1
\G^2 \G^3 \G^4$.

In computing the Killing spinor, one needs to know the self dual and
anti-self dual parts of the spin connection and field strength. For
the field strength these are
\bea
F^{ab} \s_{ab} & = & \frac{- 2 i {\bf \t} \cdot \nabla U}{U^2} \,, \nonumber
\\
F^{ab} \tilde \s_{ab} & = & \frac{- 2 i {\bf \t} \cdot \nabla
\Ub}{\Ub^2} \,.
\eea
Whilst for the spin connection we have
\bea
\w^{ab} \s_{ab} & = & \frac{-2i}{(U\Ub)^{1/2}} \left[\frac{{\bf \t} \cdot
    \nabla U e^0}{U} + \frac{({\bf \t} \times \nabla U) \cdot e}{U} \right]\,, \nonumber \\
\w^{ab} \tilde \s_{ab} & = & \frac{-2i}{(U\Ub)^{1/2}} \left[- \frac{{\bf \t} \cdot
    \nabla \Ub e^0}{\Ub} + \frac{({\bf \t} \times \nabla \Ub) \cdot e}{\Ub} \right] \,.
\eea

\section{Two component spinor conventions}

We can use the matrices of Appendix B to relate vectors in four
component notation to two component spinor notation
\be
X^{A A'} = \frac{-i \s^{A A'}_a}{\sqrt{2}} X^a \,.
\ee
Because $a$ is a Euclidean signature tangent space index, raising and
lowering this index does not have any effect. The inverse to this
relation is
\be
X^a = \frac{i \s^a_{B' B}}{\sqrt{2}} \tilde X^{B B'} \,,
\ee
which works because
\be
\s^{A A'}_a \tilde \s^a_{B' B} = 2 \delta^A_B \d^{A'}_{B'} \,.
\ee
Another useful relation is
\be
\tilde \s^a_{A' A} \tilde \s_{a \, B' B} = -2 \ep_{AB} \ep_{A' B'} \,,
\ee
and similarly for the $\s_a^{A A'}$.

\section{Equivalence and inequivalence of inner products}

Suppose a metric perturbation satisfies the gauge condition
\be\label{eq:gauge1}
\nabla^{\nu} \d g_{\mu \nu} = \lambda \nabla_{\mu} \d
  g^{\nu}{}_{\nu} \,.
\ee
Consider the same perturbation in a different gauge,
\be\label{eq:gauge2}
\nabla^{\nu} \d \hat g_{\mu \nu} = \hat \lambda \nabla_{\mu} \d
  \hat g^{\nu}{}_{\nu} \,.
\ee
The two perturbations are thus related by
\be
\d \hat g_{\mu \nu} = \d g_{\mu \nu} + \nabla_\mu \xi_\nu +
\nabla_\nu \xi_\mu \,,
\ee
and the two gauge conditions (\ref{eq:gauge1}) and
(\ref{eq:gauge2}) require that $\xi$ satisfy
\be
2 \nabla^\mu \nabla_{(\mu} \xi_{\nu)} = (\hat \lambda - \lambda)
\nabla_\nu \d g^{\mu}{}_{\mu} + 2 \hat \lambda \nabla_\nu
\nabla^\mu \xi_\mu \,.
\ee
Using this relation it is straightforward to show that
\be
\langle \d \hat g, \d \hat g \rangle_{\hat \lambda} = \langle \d g, \d g
\rangle_{\lambda} + 2(\lambda - \hat \lambda) \int d^4x \sqrt{g} \d
g^\mu{}_\mu \d \hat g^\nu{}_\nu \,.
\ee
Therefore, if $\lambda \neq \hat \lambda$ the two inner products
are generically inequivalent. They are equivalent if all the modes
are trace free in one of the gauges. This is consistent with the
observation in the main text that the inner products are
manifestly equivalent on trace free modes. We also noted in the
text that on noncompact gravitational instantons all normalisable
zero modes are indeed trace free, as follows from the fact that
these satisfy $\nabla^2 \d g^{\mu}{}_\mu = 0$. However, even in
this noncompact pure gravity case, nonzero modes will of course
generically have a trace component.

\end{document}